\documentclass{svmult}
\usepackage{epsfig,graphicx} 
\usepackage[bottom]{footmisc}
\usepackage{natbib,url}      
\bibpunct{(}{)}{;}{a}{}{,}   
\def\cite#1{\citealp{#1}}    
\def\authorindex#1{}  
\def\figspath{.}

\begin{document}\newcount\preprintheader\preprintheader=1
{ 

\def\thisvolume{these proceedings}

\def\aj{{AJ}}			
\def\araa{{ARA\&A}}		
\def\apj{{ApJ}}			
\def\apjl{{ApJ}}		
\def\apjs{{ApJS}}		
\def\ao{{Appl.\ Optics}} 
\def\apss{{Ap\&SS}}		
\def\aap{{A\&A}}		
\def\aapr{{A\&A~Rev.}}		
\def\aaps{{A\&AS}}		
\def\an{{Astron.\ Nachrichten}}
\def\aspcs{{ASP Conf.\ Ser.}}
\def\assp{{Astrophys.\ \& Space Sci.\ Procs., Springer, Heidelberg}}
\def\azh{{AZh}}			
\def\baas{{BAAS}}		
\def\jrasc{{JRASC}}	
\def\memras{{MmRAS}}		
\def\mnras{{MNRAS}}
\def\nat{{Nat}}		
\def\pra{{Phys.\ Rev.\ A}} 
\def\prb{{Phys.\ Rev.\ B}}		
\def\prc{{Phys.\ Rev.\ C}}		
\def\prd{{Phys.\ Rev.\ D}}		
\def\prl{{Phys.\ Rev.\ Lett.}} 
\def\pasp{{PASP}}
\def\pasj{{PASJ}}		
\def\qjras{{QJRAS}}
\def\science{{Sci}}		
\def\skytel{{S\&T}}		
\def\solphys{{Solar\ Phys.}} 
\def\sovast{{Soviet\ Ast.}}  
\def\ssr{{Space\ Sci.\ Rev.}}
\def\svassp{{Astrophys.\ Space Sci.\ Procs., Springer, Heidelberg}}
\def\zap{{ZAp}}			
\let\astap=\aap
\let\apjlett=\apjl
\let\apjsupp=\apjs
\def\grl{{Geophys.\ Res.\ Lett.}}  
\def\jgr{{J. Geophys.\ Res.}} 

\def\ion#1#2{{\rm #1}\,{\uppercase{#2}}}  
\def\deg{\hbox{$^\circ$}}
\def\sun{\hbox{$\odot$}}
\def\earth{\hbox{$\oplus$}}
\def\la{\mathrel{\hbox{\rlap{\hbox{\lower4pt\hbox{$\sim$}}}\hbox{$<$}}}}
\def\ga{\mathrel{\hbox{\rlap{\hbox{\lower4pt\hbox{$\sim$}}}\hbox{$>$}}}}
\def\sq{\hbox{\rlap{$\sqcap$}$\sqcup$}}
\def\arcmin{\hbox{$^\prime$}}
\def\arcsec{\hbox{$^{\prime\prime}$}}
\def\fd{\hbox{$.\!\!^{\rm d}$}}
\def\fh{\hbox{$.\!\!^{\rm h}$}}
\def\fm{\hbox{$.\!\!^{\rm m}$}}
\def\fs{\hbox{$.\!\!^{\rm s}$}}
\def\fdg{\hbox{$.\!\!^\circ$}}
\def\farcm{\hbox{$.\mkern-4mu^\prime$}}
\def\farcs{\hbox{$.\!\!^{\prime\prime}$}}
\def\fp{\hbox{$.\!\!^{\scriptscriptstyle\rm p}$}}
\def\micron{\hbox{$\mu$m}}
\def\onehalf{\hbox{$\,^1\!/_2$}}	
\def\onethird{\hbox{$\,^1\!/_3$}}
\def\twothirds{\hbox{$\,^2\!/_3$}}
\def\onequarter{\hbox{$\,^1\!/_4$}}
\def\threequarters{\hbox{$\,^3\!/_4$}}
\def\ubv{\hbox{$U\!BV$}}		
\def\ubvr{\hbox{$U\!BV\!R$}}		
\def\ubvri{\hbox{$U\!BV\!RI$}}		
\def\ubvrij{\hbox{$U\!BV\!RI\!J$}}		
\def\ubvrijh{\hbox{$U\!BV\!RI\!J\!H$}}		
\def\ubvrijhk{\hbox{$U\!BV\!RI\!J\!H\!K$}}		
\def\ub{\hbox{$U\!-\!B$}}		
\def\bv{\hbox{$B\!-\!V$}}		
\def\vr{\hbox{$V\!-\!R$}}		
\def\ur{\hbox{$U\!-\!R$}}


\def\labelitemi{{\bf --}}  

\def\rmit#1{{\it #1}}              
\def\rmit#1{{\rm #1}}              
\def\etal{\rmit{et al.}}           
\def\etc{\rmit{etc.}}           
\def\ie{\rmit{i.e.,}}              
\def\eg{\rmit{e.g.,}}              
\def\cf{cf.}                       
\def\viz{\rmit{viz.}}
\def\vs{\rmit{vs.}}

\def\rot{\hbox{\rm rot}}
\def\div{\hbox{\rm div}}
\def\lesssim{\mathrel{\hbox{\rlap{\hbox{\lower4pt\hbox{$\sim$}}}\hbox{$<$}}}}
\def\gtrsim{\mathrel{\hbox{\rlap{\hbox{\lower4pt\hbox{$\sim$}}}\hbox{$>$}}}}
\def\mathstacksym#1#2#3#4#5{\def#1{\mathrel{\hbox to 0pt{\lower 
    #5\hbox{#3}\hss} \raise #4\hbox{#2}}}}
\mathstacksym\lesssim{$<$}{$\sim$}{1.5pt}{3.5pt} 
\mathstacksym\gtrsim{$>$}{$\sim$}{1.5pt}{3.5pt} 
\mathstacksym\lrarrow{$\leftarrow$}{$\rightarrow$}{2pt}{1pt} 
\mathstacksym\lessgreat{$>$}{$<$}{3pt}{3pt} 

\def\dif{\: {\rm d}}                       
\def\ep{\:{\rm e}^}                        
\def\dash{\hbox{$\,-\,$}}                  
\def\is{\!=\!}                             

\def\starname#1#2{${#1}$\,{\rm {#2}}}  
\def\Teff{\hbox{$T_{\rm eff}$}}   

\def\kms{\hbox{km$\;$s$^{-1}$}}
\def\ms{\hbox{m$\;$s$^{-1}$}}
\def\Mxcm{\hbox{Mx\,cm$^{-2}$}}    

\def\Bapp{\hbox{$B_{\rm app}$}}    

\def\komega{($k, \omega$)}                 
\def\kf{($k_h,f$)}                         
\def\VminI{\hbox{$V\!\!-\!\!I$}}           
\def\IminI{\hbox{$I\!\!-\!\!I$}}           
\def\VminV{\hbox{$V\!\!-\!\!V$}}           
\def\Xt{\hbox{$X\!\!-\!t$}}                

\def\level #1 #2#3#4{$#1 \: ^{#2} \mbox{#3} ^{#4}$}   

\def\specchar#1{\uppercase{#1}}    
\def\AlI{\mbox{Al\,\specchar{i}}}  
\def\BI{\mbox{B\,\specchar{i}}} 
\def\BII{\mbox{B\,\specchar{ii}}}  
\def\BaI{\mbox{Ba\,\specchar{i}}}  
\def\BaII{\mbox{Ba\,\specchar{ii}}} 
\def\CI{\mbox{C\,\specchar{i}}} 
\def\CII{\mbox{C\,\specchar{ii}}} 
\def\CIII{\mbox{C\,\specchar{iii}}} 
\def\CIV{\mbox{C\,\specchar{iv}}} 
\def\CaI{\mbox{Ca\,\specchar{i}}} 
\def\CaII{\mbox{Ca\,\specchar{ii}}} 
\def\CaIII{\mbox{Ca\,\specchar{iii}}} 
\def\CoI{\mbox{Co\,\specchar{i}}} 
\def\CrI{\mbox{Cr\,\specchar{i}}} 
\def\CriI{\mbox{Cr\,\specchar{ii}}} 
\def\CsI{\mbox{Cs\,\specchar{i}}} 
\def\CsII{\mbox{Cs\,\specchar{ii}}} 
\def\CuI{\mbox{Cu\,\specchar{i}}} 
\def\FeI{\mbox{Fe\,\specchar{i}}} 
\def\FeII{\mbox{Fe\,\specchar{ii}}} 
\def\FeIX{\mbox{Fe\,\specchar{ix}}}
\def\FeX{\mbox{Fe\,\specchar{x}}}
\def\FeXVI{\mbox{Fe\,\specchar{xvi}}}
\def\FrI{\mbox{Fr\,\specchar{i}}}
\def\HI{\mbox{H\,\specchar{i}}} 
\def\HII{\mbox{H\,\specchar{ii}}} 
\def\Hmin{\hbox{\rmH$^{^{_{\scriptstyle -}}}$}}      
\def\Hemin{\hbox{{\rm He}$^{^{_{\scriptstyle -}}}$}} 
\def\HeI{\mbox{He\,\specchar{i}}} 
\def\HeII{\mbox{He\,\specchar{ii}}} 
\def\HeIII{\mbox{He\,\specchar{iii}}} 
\def\KI{\mbox{K\,\specchar{i}}} 
\def\KII{\mbox{K\,\specchar{ii}}} 
\def\KIII{\mbox{K\,\specchar{iii}}} 
\def\LiI{\mbox{Li\,\specchar{i}}} 
\def\LiII{\mbox{Li\,\specchar{ii}}} 
\def\LiIII{\mbox{Li\,\specchar{iii}}} 
\def\MgI{\mbox{Mg\,\specchar{i}}} 
\def\MgII{\mbox{Mg\,\specchar{ii}}} 
\def\MgIII{\mbox{Mg\,\specchar{iii}}} 
\def\MnI{\mbox{Mn\,\specchar{i}}} 
\def\NI{\mbox{N\,\specchar{i}}}
\def\NIV{\mbox{N\,\specchar{iv}}}
\def\NaI{\mbox{Na\,\specchar{i}}}
\def\NaII{\mbox{Na\,\specchar{ii}}}
\def\NaIII{\mbox{Na\,\specchar{iii}}}
\def\NeVIII{\mbox{Ne\,\specchar{viii}}} 
\def\NiI{\mbox{Ni\,\specchar{i}}} 
\def\NiII{\mbox{Ni\,\specchar{ii}}}
\def\NiIII{\mbox{Ni\,\specchar{iii}}} 
\def\OI{\mbox{O\,\specchar{i}}} 
\def\OVI{\mbox{O\,\specchar{vi}}}
\def\RbI{\mbox{Rb\,\specchar{i}}} 
\def\SII{\mbox{S\,\specchar{ii}}} 
\def\SiI{\mbox{Si\,\specchar{i}}} 
\def\SiII{\mbox{Si\,\specchar{ii}}} 
\def\SrI{\mbox{Sr\,\specchar{i}}}
\def\SrII{\mbox{Sr\,\specchar{ii}}}
\def\TiI{\mbox{Ti\,\specchar{i}}} 
\def\TiII{\mbox{Ti\,\specchar{ii}}} 
\def\TiIII{\mbox{Ti\,\specchar{iii}}} 
\def\TiIV{\mbox{Ti\,\specchar{iv}}} 
\def\VI{\mbox{V\,\specchar{i}}} 
\def\HtwoO{\mbox{H$_2$O}}        
\def\Otwo{\mbox{O$_2$}}          

\def\Halpha{\mbox{H\hspace{0.1ex}$\alpha$}} 
\def\Ha{\mbox{H\hspace{0.2ex}$\alpha$}}
\def\Hbeta{\mbox{H\hspace{0.2ex}$\beta$}}
\def\Hgamma{\mbox{H\hspace{0.2ex}$\gamma$}}
\def\Hdelta{\mbox{H\hspace{0.2ex}$\delta$}}
\def\Hepsilon{\mbox{H\hspace{0.2ex}$\epsilon$}}
\def\Hzeta{\mbox{H\hspace{0.2ex}$\zeta$}}
\def\Lyalpha{\mbox{Ly$\hspace{0.2ex}\alpha$}}
\def\Lybeta{\mbox{Ly$\hspace{0.2ex}\beta$}}
\def\Lygamma{\mbox{Ly$\hspace{0.2ex}\gamma$}}
\def\Lycont{\mbox{Ly\hspace{0.2ex}{\small cont}}}
\def\Baalpha{\mbox{Ba$\hspace{0.2ex}\alpha$}}
\def\Babeta{\mbox{Ba$\hspace{0.2ex}\beta$}}
\def\Bacont{\mbox{Ba\hspace{0.2ex}{\small cont}}}
\def\Paalpha{\mbox{Pa$\hspace{0.2ex}\alpha$}}
\def\Bralpha{\mbox{Br$\hspace{0.2ex}\alpha$}}

\def\NaD{\mbox{Na\,\specchar{i}\,D}}    
\def\NaDone{\mbox{Na\,\specchar{i}\,\,D$_1$}}
\def\NaDtwo{\mbox{Na\,\specchar{i}\,\,D$_2$}}
\def\NaID{\mbox{Na\,\specchar{i}\,\,D}}
\def\NaIDone{\mbox{Na\,\specchar{i}\,\,D$_1$}}
\def\NaIDtwo{\mbox{Na\,\specchar{i}\,\,D$_2$}}
\def\Done{\mbox{D$_1$}}
\def\Dtwo{\mbox{D$_2$}}

\def\Mgbone{\mbox{Mg\,\specchar{i}\,b$_1$}}
\def\Mgbtwo{\mbox{Mg\,\specchar{i}\,b$_2$}}
\def\Mgbthree{\mbox{Mg\,\specchar{i}\,b$_3$}}
\def\MgIb{\mbox{Mg\,\specchar{i}\,b}}
\def\MgIbone{\mbox{Mg\,\specchar{i}\,b$_1$}}
\def\MgIbtwo{\mbox{Mg\,\specchar{i}\,b$_2$}}
\def\MgIbthree{\mbox{Mg\,\specchar{i}\,b$_3$}}

\def\CaIIK{\mbox{Ca\,\specchar{ii}\,K}}       
\def\CaIIH{\mbox{Ca\,\specchar{ii}\,H}}
\def\CaIIHK{\mbox{Ca\,\specchar{ii}\,H\,\&\,K}}
\def\HK{\mbox{H\,\&\,K}}
\def\Kthree{\mbox{K$_3$}}      
\def\Hthree{\mbox{H$_3$}}
\def\Ktwo{\mbox{K$_2$}}
\def\Htwo{\mbox{H$_2$}}
\def\Kone{\mbox{K$_1$}}     
\def\Hone{\mbox{H$_1$}}     
\def\KtwoV{\mbox{K$_{2V}$}}
\def\KtwoR{\mbox{K$_{2R}$}}
\def\KoneV{\mbox{K$_{1V}$}}
\def\KoneR{\mbox{K$_{1R}$}}
\def\HtwoV{\mbox{H$_{2V}$}}
\def\HtwoR{\mbox{H$_{2R}$}}
\def\HoneV{\mbox{H$_{1V}$}}
\def\HoneR{\mbox{H$_{1R}$}}

\def\hk{\mbox{h\,\&\,k}}
\def\kthree{\mbox{k$_3$}}    
\def\hthree{\mbox{h$_3$}}
\def\ktwo{\mbox{k$_2$}}
\def\htwo{\mbox{h$_2$}}
\def\kone{\mbox{k$_1$}}     
\def\hone{\mbox{h$_1$}}     
\def\ktwoV{\mbox{k$_{2V}$}}
\def\ktwoR{\mbox{k$_{2R}$}}
\def\koneV{\mbox{k$_{1V}$}}
\def\koneR{\mbox{k$_{1R}$}}
\def\htwoV{\mbox{h$_{2V}$}}
\def\htwoR{\mbox{h$_{2R}$}}
\def\honeV{\mbox{h$_{1V}$}}
\def\honeR{\mbox{h$_{1R}$}}

\ifnum\preprintheader=1     
\makeatletter  
\def\@maketitle{\newpage
\markboth{}{}%
  {\mbox{}
  \vspace*{-8ex} \par 
  \em \footnotesize To appear in ``Recent Advances in Spectroscopy:
  Astrophysical, Theoretical and Experimental Perspectives'', eds.\
  R.~K.~Chaudhuri, M.~V.~Mekkaden, A.~V.~Raveendran and A.~Satya
  Narayanan, Astrophysics and Space Science Proceedings,
  Springer-Verlag, Heidelberg, Berlin, 2009.}  \vspace*{-5ex} \par
 \def\lastand{\ifnum\value{@inst}=2\relax
                 \unskip{} \andname\
              \else
                 \unskip \lastandname\
              \fi}%
 \def\and{\stepcounter{@auth}\relax
          \ifnum\value{@auth}=\value{@inst}%
             \lastand
          \else
             \unskip,
          \fi}%
  \raggedright
 {\Large \bfseries\boldmath
  \pretolerance=10000
  \let\\=\newline
  \raggedright
  \hyphenpenalty \@M
  \interlinepenalty \@M
  \if@numart
     \chap@hangfrom{}
  \else
     \chap@hangfrom{\thechapter\thechapterend\hskip\betweenumberspace}
  \fi
  \ignorespaces
  \@title \par}\vskip .8cm
\if!\@subtitle!\else {\large \bfseries\boldmath
  \vskip -.65cm
  \pretolerance=10000
  \@subtitle \par}\vskip .8cm\fi
 \setbox0=\vbox{\setcounter{@auth}{1}\def\and{\stepcounter{@auth}}%
 \def\thanks##1{}\@author}%
 \global\value{@inst}=\value{@auth}%
 \global\value{auco}=\value{@auth}%
 \setcounter{@auth}{1}%
{\lineskip .5em
\noindent\ignorespaces
\@author\vskip.35cm}
 {\small\institutename\par}
 \ifdim\pagetotal>157\p@
     \vskip 11\p@
 \else
     \@tempdima=168\p@\advance\@tempdima by-\pagetotal
     \vskip\@tempdima
 \fi
}
\makeatother     
\fi

\title*{Solar Spectroscopy and (Pseudo-)Diagnostics of the Solar
  Chromosphere}
\titlerunning{Solar Spectroscopy and the Solar Chromosphere}

\author{Robert J. Rutten}
\authorindex{Rutten, R. J.}

\institute{Sterrekundig Instituut, Utrecht University, Utrecht,
  The Netherlands\\
  Institutt for Teoretisk Astrofysikk, Oslo University, Oslo, Norway}

\maketitle

\setcounter{footnote}{0}  

\begin{abstract} 
  I first review trends in current solar spectrometry and then
  concentrate on comparing various spectroscopic diagnostics of the
  solar chromosphere.  Some are actually not at all chromospheric but
  just photospheric or clapotispheric and do not convey information on
  chromospheric heating, even though this is often assumed.  Balmer
  \Halpha\ is the principal displayer of the closed-field
  chromosphere, but it is unclear how chromospheric fibrils gain their
  large \Halpha\ opacity.  The open-field chromosphere seems to harbor
  most if not all coronal heating and solar wind driving, but is
  hardly seen in optical diagnostics.
\end{abstract}

\section{Solar spectroscopy}
During this wide-ranging conference I gave a wide-ranging review
covering spectroscopy of the solar photosphere, chromosphere, and
corona including condensed tutorials\footnote{
  \url{http://www.astro.uu.nl/~rutten/Lectures_on.html}; for a
  non-condensed syllabus see
  \url{http://www.astro.uu.nl/~rutten/Lecture_notes.html}.}  of line
formation theory for these disparate domains.  Being page-limited, I
only summarize some major review points here and then concentrate on
optical diagnostics of the solar chromosphere -- arguably the hardest
solar nut to crack (\cite{1998SSRv...85..187J}).

\subparagraph{Solar atmosphere physics.}  A major quest of solar
physics is to understand the intricate interplay between solar
magnetism, gas dynamics, and radiation that makes the solar
atmosphere such an outstanding research laboratory and also
governs our space environment.  The photosphere, chromosphere, and
corona represent optically thick, effectively thin, and optically thin
interaction domains with much variation in plasma-$\beta$.  They
differ much in research techniques and in researcher specialisms, but
do require holistic synthesis.

Spectroscopy is a key research tool in this endeavor.  It differs
from other astronomy in that the solar atmosphere displays its
interactions in resolvable fine structure that varies on short time
scales.  Solar spectroscopy therefore requires high time resolution, both
observationally and interpretationally, in addition to high angular
resolution and large diagnostic diversity.

\subparagraph{Solar physics now.}  Solar physics is experiencing a
tremendous boost from three developments: wavefront correction
enabling 0.1\arcsec\ resolution from meter-class optical telescopes,
continuous multi-wavelength high-cadence monitoring from space, and
increasing realism of numerical MHD simulations of solar-atmosphere
fine structure.  Photosphere--chromosphere--corona coupling is a key
research topic and becomes addressable at the level of detailed
understanding rather than wishful cartoon/mechanism thinking.

\subparagraph{Classical spectroscopy.}  Traditionally, solar
spectroscopy concentrated on abundance determination using solar
spectrum atlases 
and assuming LTE with a best-fit plane-parallel
hydrostatic-equilibrium model photosphere (review:
\cite{2002JAD.....8....8R}). 
This practice and the resulting standards in helioseismology and
stellar evolution theory have been upset by Asplund's turning to 3D
time-dependent Nordlund-Stein granulation simulations and obtaining
significantly smaller abundance values for key elements such as
oxygen, carbon and nitrogen
(\eg\ \cite{2004A&A...417..751A}; 
review: \cite{2009LRSP....6....2N}). 

\subparagraph{Trends in observation.}  The emphasis of current solar
spectrometry is on imaging spectroscopy with high image quality,
improving spatial and temporal coverage and resolution by limiting the
spectral sampling to only moderate spectral resolution in only
specific diagnostic lines.  Optically this is done best with
Fabry-P\'erot interferometers (IBIS,
\cite{2006SoPh..236..415C}, 
\cite{2008A&A...481..897R}; 
CRISP, \cite{2006A&A...447.1111S}, 
\cite{2008ApJ...689L..69S}; 
the new G\"ottingen one,
\cite{2008A&A...480..265B}). 
They sample selected line profiles sequentially in time.  Instantaneous
integral-field spectrometry using fiber and/or lenslet arrays for
field reformatting is highly desirable but remains on the horizon
(\cite{Rutten1999c}; 
\cite{2004ApJ...613L.177L}; 
\cite{Sankar++2009}). 
Slit spectrometers are still in use, but only for precision
spectropolarimetry and for ultraviolet emission-line measurement from
space, in both cases for want of suited integral-field technology.
Slits are no good because they always sample the wrong place at the
wrong time.  In ground-based observation, they also inhibit wavefront
restoration by algorithms such as the very effective MOMFBD of
\citet{2005SoPh..228..191V}, 
a must in addition to adaptive optics.  In space-based observation,
multi-layer mirror technology has enabled the fruitful narrow-passband EUV
imaging of SOHO, TRACE, STEREO and the comprehensive monitoring
promised by SDO, but, unfortunately, ultraviolet integral-field
spectrometry remains a too distant hope.

\subparagraph{Trends in interpretation.}  Solar-atmosphere radiative
transfer modeling is presently split between data inversions and
simulatory forward modeling.  Inversion codes initiated by Ruiz-Cobo
et al.\ (1990, 1992) \nocite{1990Ap&SS.170..113R} 
\nocite{1992ApJ...398..375R} 
replace the classical ``semi-empirical'' (= best-fit)
``one-dimensional'' (= plane-parallel hydrostatic-equilibrium)
modeling that either assumed LTE and CRD
(\cite{1967ZA.....65..365H}; 
revised into HOLMUL by \cite{1974SoPh...39...19H}) 
or relaxed these assumptions to NLTE and PRD (VALC,
\cite{1981ApJS...45..635V}, 
revised into FALC by
\cite{1993ApJ...406..319F}). 
The data inversions obtain such best-fit stratifications per pixel on
the solar surface and are applied particularly in photospheric
spectropolarimetry.  Just as for the standard models, they loose
credibility higher up where clapotispheric (see below) and
chromospheric fluctuations are too wild for smooth spline-function
fitting or temporal averaging and vertical columns too poor a sample
of 3D structure. Chromospheric line formation requires forward
modeling in the form of NLTE line synthesis within numerical
simulations including non-equilibrium ionization evaluation (see
below).  A major task is to implement 3D time-dependent radiative
transfer in 3D time-dependent MHD codes.  A major challenge is to make
such codes fast.

\begin{figure}  
 \sidecaption
  \includegraphics[width=80mm]{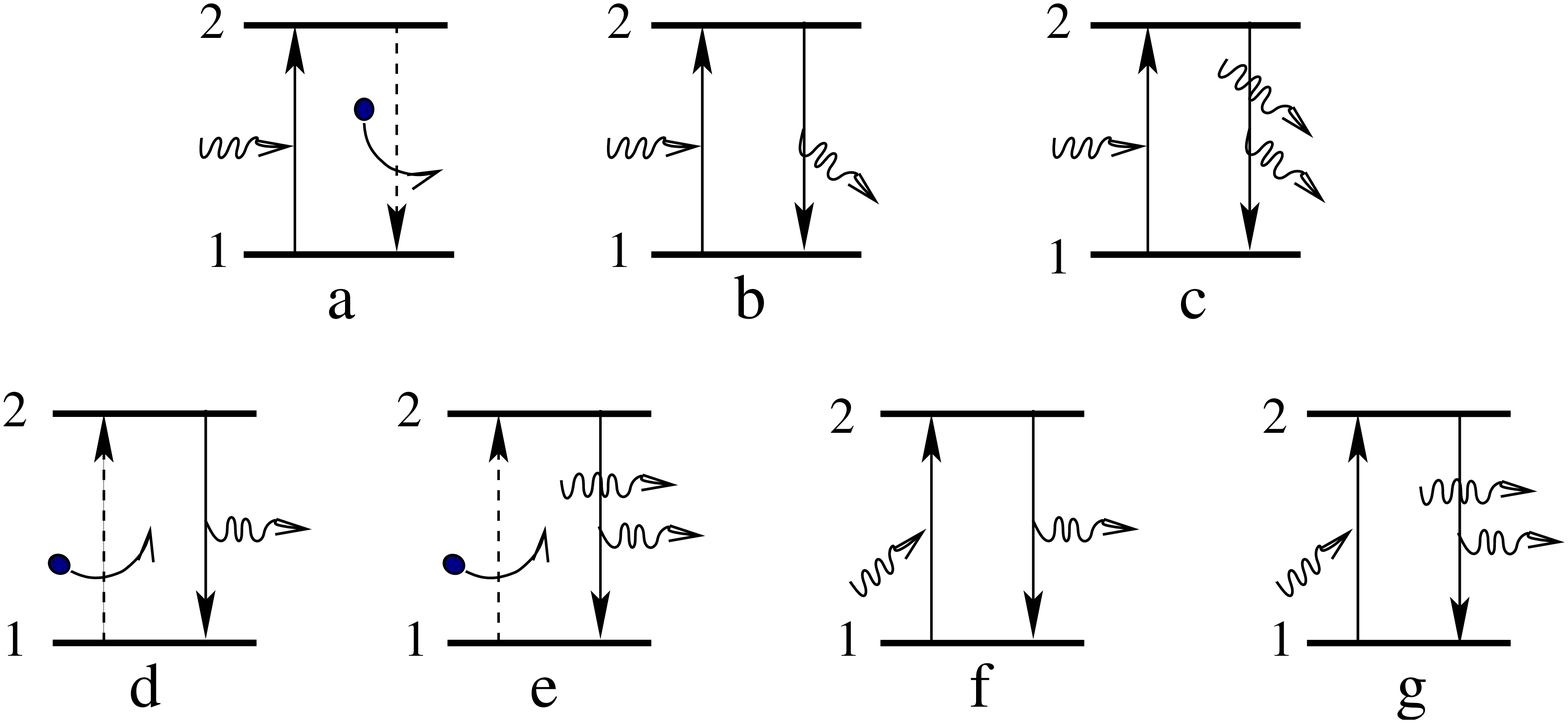}
  \caption[]{\label{rutten-fig:2levelpairs} %
    Bound-bound process pairs contributing line extinction and
    emissivity.  {\em Pair a\/}: photon destruction.  {\em Pairs b and
      c\/}: spontaneous and induced scattering out of the beam.  {\em
      Pairs d and e\/}: photon creation.  {\em Pairs f and g\/}:
    scattering into the beam.  From
    \citet{Rutten2003:RTSA}. 
  }
\end{figure}

\section{Solar line formation}
Fig.~\ref{rutten-fig:2levelpairs} displays all radiative
excitation--deexcitation process combinations for idealized two-level
atoms.  Their competition defines different radiation-physics domains
of spectral line formation:
  \begin{itemize} \itemsep=1ex
   \item {\em LTE = large collision frequency} 
     \begin{itemize} 
        \item up: mostly collisional = thermal creation (d + e);
        \item down: mostly collisional = large destruction probability (a);
        \item photon travel: ``honorary gas particles'' or negligible leak.
     \end{itemize}
   \item {\em NLTE = statistical equilibrium or time-dependent}
     \begin{itemize} 
     \item up and down: mostly radiative = non-local sources;
        \item two-level scattering: 
              coherent/complete/partial redistribution;
            \item multi-level travel: pumping, suction, sensitivity
              transcription.
     \end{itemize}
   \item {\em coronal equilibrium = hot tenuous}
     \begin{itemize} 
        \item up: only collisional = thermal creation (only d);
        \item down: only spontaneous (only d);
        \item photon travel: escape / drown / scatter bf H\,I, He\,I, He\,II.
     \end{itemize}
\end{itemize}
that also characterize line formation in solar-atmosphere domains:
\begin{itemize} \itemsep=1ex

\item{\em Photosphere.}  Optical continuum and weak-line photon
  escape.  High density makes LTE (= Saha--Boltzmann population ratios
  producing $S_\nu = B_\nu$) valid for optical continua and reasonable
  for most subordinate lines.  ``Height of formation'' is a valid
  concept.  The major trick is the Eddington-Barbier
  approximation $I_\nu(\tau_\nu=0, \mu) \approx S_\nu(\tau_\nu=\mu)
  \approx B_\nu[T(\tau_\nu=\mu)]$ for quick insight and 
  inversion constraining.  Upper-photosphere resonance lines such
  as the \NaID\ lines, \KI~7699\,\AA, and \BaII~4554\,\AA\ darken from
  simple resonance scattering
  (\cite{1992A&A...265..268U}). 

\item{\em Chromosphere.}  Strong-line photon escape.  LTE is invalid,
  statistical equilibrium is invalid, complete redistribution is
  invalid, instantaneous ionization equilibrium is invalid,
  single-fluid description is likely to be invalid.  Height of
  formation: fibrils somewhere between the photosphere and the
  telescope (\cite{2007ASPC..368...27R}). 
  The major trick is the four-parameter ($S_\nu$, $\tau_\nu$,
  $\lambda-\lambda_0$, $\Delta \lambda_{\rm D}$) cloud model of
  \citet{Beckers1964}, 
  but all applications deriving temperature and density using \Halpha\
  have erred in assuming instantaneous ionization/recombination
  balancing (see below).

\item{\em Corona.}  Immediate photon escape.  Optical thinness plus
  the (d)-only equilibrium (but adding dielectronic
  recombination into the bound-free analogon) give the 
  major trick: $ \sum h\nu
  \propto \int n_{\rm ion} N_{\rm e} \, \dif z = \int N_{\rm e}^2
  \left( {\rm d} T/{\rm d} z \right)^{-1} \dif T \equiv $ ``emission
  measure''.  Resonance scattering may affect the strongest lines.
  Bound-free scattering out of EUV-line filtergraph passbands makes
  neutral-hydrogen clouds appear thickly black against bright EUV-line
  backgrounds in EUV filtergrams
  (see the cartoon in Fig.~10 of \cite{Rutten1999d}). 

\end{itemize}

\section{Clapotisphere versus chromosphere}
The LTE line-fitting HOLMUL model favored by pre-Asplund abundance
determiners has no chromosphere whatsoever but a temperature
stratification close to radiative equilibrium to avoid non-observed
reversals (emission cores) in strong \FeI\ lines.  Admitting realistic
NLTE departures delivers the same apparent line-core excitation
temperatures when there is a chromospheric temperature rise above $h
\approx 500$~km (\cite{Rutten+Kostik1982}), 
and therefore the NLTE VALC/FALC models fudged the ultraviolet
line haze (\cite{1975SoPh...45..115Z}) 
to resemble HOLMUL LTE lines by imposing a gradual transition from pure
absorption to pure scattering over the upper photosphere
(\cite{Rutten1988b}), 
so that they don't feel the VALC/FALC temperature rise which became
known as ``the chromosphere''.  However, 
\citet{1997ApJ...481..500C} 
showed that such a global temperature rise is unlikely to occur below
$h \approx 1000$~km in quiet-Sun internetwork, where intermittent
shocks momentarily raise the temperature in brief spikes that dominate
the time-averaged ultraviolet emissivity.  I called this shock-ridden
regime the ``clapotisphere'' 
(\cite{Rutten1995b}) 
using a nautical term for fierce wave interaction.  It is cartoonized
in Fig.~\ref{rutten-fig:potsdam-f2}.

\begin{figure}  
  \centering    
  \includegraphics[width=0.8\textwidth]{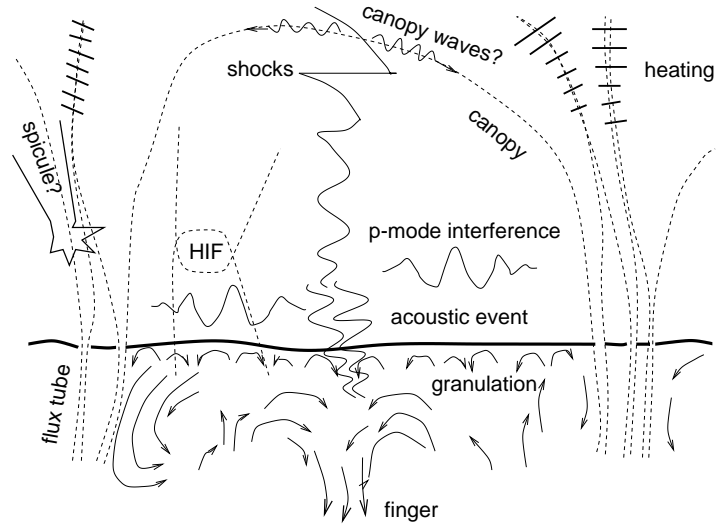}
  \caption[]{\label{rutten-fig:potsdam-f2} %
    Cartoon of the sub-canopy clapotisphere, pervaded by three-minute
    shocks.  HIF stands for horizontal internetwork field
    (\cite{1996ApJ...460.1019L}). 
    From \citet{Rutten1999d}. 
}\end{figure}

\begin{figure}  
  \centering    
  \includegraphics[width=\textwidth]{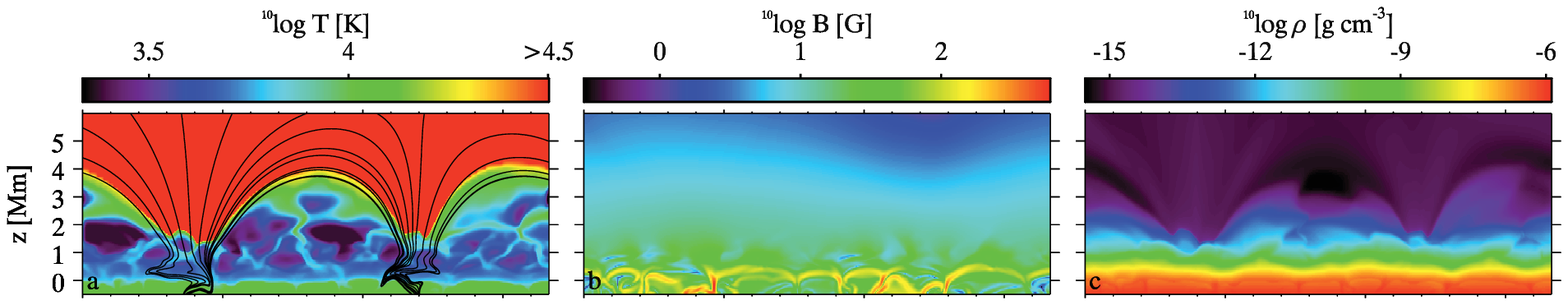}
  \includegraphics[width=\textwidth]{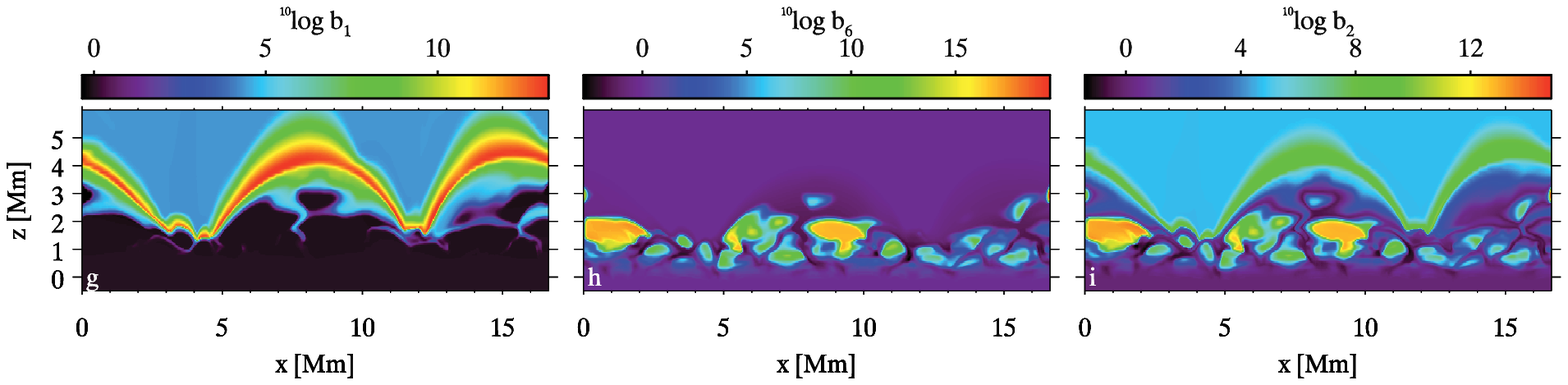}
  \caption[]{\label{rutten-fig:hion-f1} %
    Snapshot from a 2D MHD Oslo STAGGER simulation with
    non-equilibrium hydrogen ionization.  A movie version is available
    at
    \url{http://www.astro.uu.nl/~rutten/rrweb/talk-material/hion2_fig1_movie.mov}.
    Each panel is a vertical plane reaching from the top of the
    convection zone to the corona.  {\em Upper row\/}: temperature
    with selected magnetic field lines, field strength, gas density.
    {\em Lower row\/}: NLTE population departure coefficients for the
    ground state, ion state, and first excited state of hydrogen.  All
    color coding is logarithmic.  Two fluxtube-like magnetic
    concentrations are connected by high-arching field that is
    outlined by very large hydrogen ground-state and excited-state
    overpopulations (red and green arches in the lower row). The
    internetwork-like domain under the arches is mostly cool but
    pervaded by interacting shocks (clouds in the last two panels).  The
    movie shows dynamic fibrils
    (\cite{2007ApJ...655..624D}) 
    that jut out from the magnetic concentrations and expand and
    contract in rapid succession.  From
    \citet{2007A&A...473..625L}. 
  }\end{figure}

\begin{figure}  
  \centering    
  \includegraphics[width=0.49\textwidth]{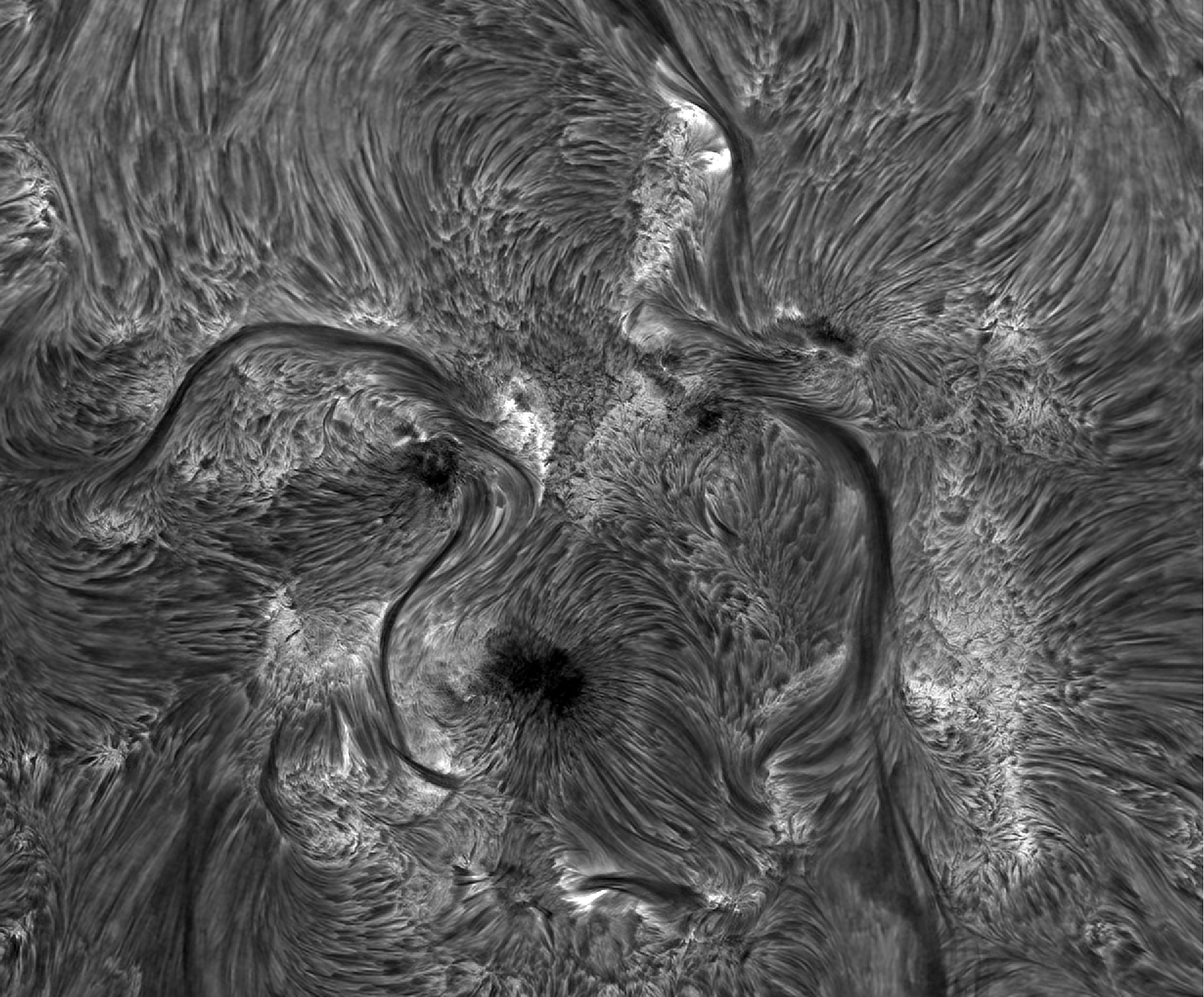}
  \includegraphics[width=0.49\textwidth]{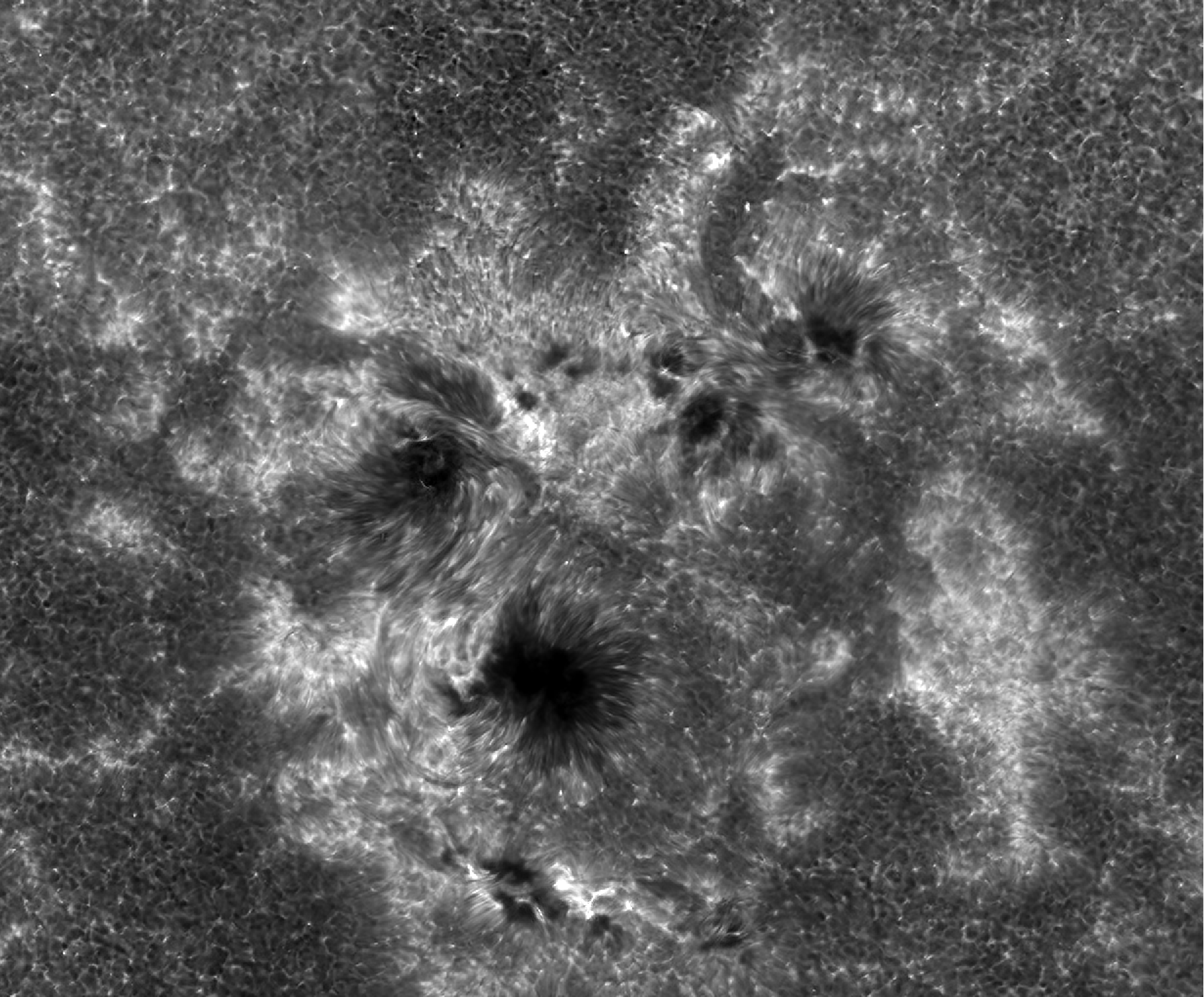}
  \caption[]{\label{rutten-fig:mosaics} %
    Active region AR10786 imaged synchronously in \Halpha\ and \CaIIH\
    by P.~S\"utterlin with the Dutch Open Telescope (DOT) on La Palma
    on July 8, 2005.  Field size $182 \times 133$ arcsec.  The
    \Halpha\ imaging used a tunable Lyot filter with passband
    0.25~\AA\ FWHM, the \CaIIH\ imaging an interference filter with
    passband 1.35~\AA\ FWHM.  The DOT also takes images synchronously
    in the G band, blue and red continua, and since recently in the
    Ba\,II\,455\,\AA\ line.  Many more DOT images, many DOT movies,
    and all reduced DOT data are available at
    \url{http://dot.astro.uu.nl}.  }\end{figure}

\begin{figure}  
  \centering    
  \includegraphics[width=1.0\textwidth]{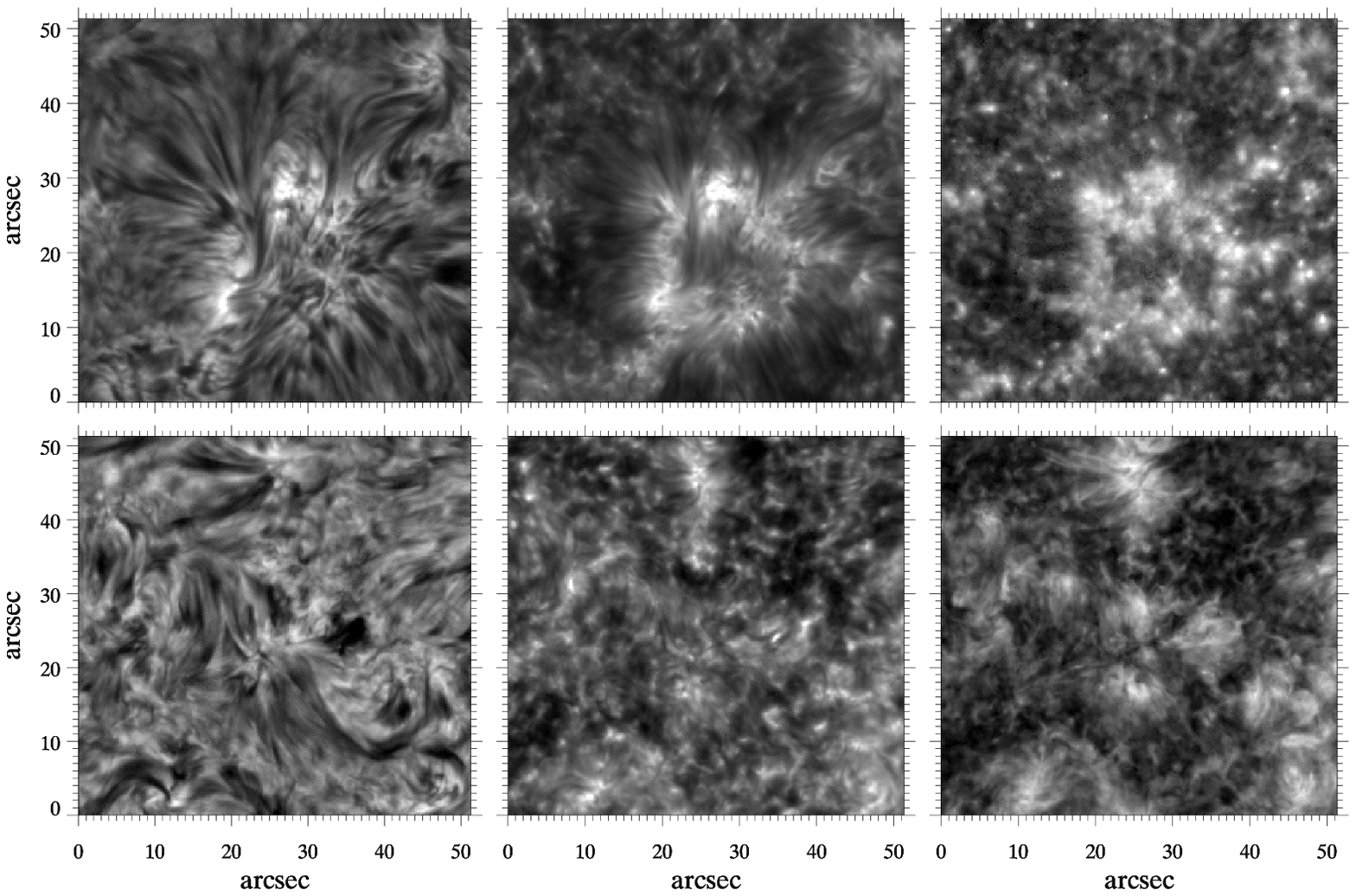}
  \caption[]{\label{rutten-fig:na-ca-ha} %
    Two sets of IBIS images.  {\em Upper row\/}: active network in
    \Halpha, \CaII\ 8542\,\AA, \NaDone.  {\em Lower row\/}: quiet
    network in \Halpha, \CaII\ 8542\,\AA, \Halpha\ core width.  The first
    five panels display the intensity of the line-profile minimum per
    pixel, the last panel the FWHM of the chromospheric \Halpha\ core
    per pixel.  IBIS is the Italian double Fabry-P\'erot imaging
    spectrometer at the Dunn Solar Telescope of the US National Solar
    Observatory.  More detail in \citet{Cauzzi++2009}. 
    Courtesy Kevin Reardon.  }\end{figure}

Such cartoons are now being vindicated by the numerical STAGGER
simulations performed in Oslo
(\cite{2007ASPC..368..107H}). 
The STAGGER snapshot in Fig.~\ref{rutten-fig:hion-f1} shows smooth
field at heights above the canopies outlined by high arches of
exceedingly large internetwork hydrogen $n\is1$ and $n\is2$ NLTE
overpopulations.  Clapotispheric shocks interact and dominate beneath
these arches.  The intershock phases are cool (first panel) and have
large ion and $n\is2$ overpopulations (last two panels).  These arise
from sluggish recombination.  Hydrogen ionizes fast in the hot shocks,
but does not recombine quickly enough in the cool aftermath to settle
its ion/atom balancing before the next shock comes along.  The
overpopulation of the $n\is2$ level, which sets the opacity of
\Halpha, follows the ground-state overpopulations in the arches and the
ion overpopulations in the clapotisphere.

So what is the chromosphere?  Classically, the name denotes the thin
\Halpha-dominated pink crescents around the eclipsed Sun glimpsed just
after second and just before third contact\footnote{As on July 22 this
  year here in India if the weather suits.}.  The \Halpha\ image in
Fig.~\ref{rutten-fig:mosaics} shows that this pink emission is
dominated by long fibrils that connect various parts of active
regions.  In quieter areas, such \Halpha\ fibrils typically connect
network across cell interiors.  These fibrils constitute the
chromosphere.  The parallel \CaIIH\ image maps them only in the
active region and only partially even there.  Elsewhere it displays
clapotispheric sub-canopy shocks and \HtwoV\ grains.  Thus, the
\Halpha\ image shows the chromosphere whereas the \CaIIH\ image is
partly clapotispheric and partly chromospheric (but mostly pseudo, see
below).

Fig.~\ref{rutten-fig:na-ca-ha} shows similar comparisons.  The active
network again shows long network-spanning fibrils in \Halpha, but the
quiet region has the clapotisphere poking through away from network
even in \Halpha\ (\cf\
\cite{2007ApJ...660L.169R}; 
\cite{2008SoPh..251..533R}). 
The upper \CaII\ 8542\,\AA\ image shows the fibrils partially, the
lower one is mostly clapotispheric away from the network.  The
\NaDone\ image (third panel) also, or not even clapotispheric
but just upper-photospheric since the IBIS spectroscopy shows
virtually no shock signatures (Kevin Reardon, private communication).
The final panel is discussed below.

\section{Pseudo diagnostics of the chromosphere}
Many spectral features that are commonly supposed to be chromospheric
actually are ``below-the-surface viewers'' for which enhanced
brightness has nothing to do with chromospheric emissivity or heating
but with deep photon escape.  The viewing pipes are Spruit fluxtubes,
slender kilogauss magnetic concentrations that occur only sparsely and
intermittently in quiet-Sun supergranulation cell interiors
(\cite{2005A&A...441.1183D}) 
but assemble more stably in supergranular cell boundaries to form the
magnetic network.  At yet larger density they constitute plage.
Fig.~\ref{rutten-fig:BPcartoon} explains how they appear as ``bright
points'' near disk center and as ``faculae'' towards the limb.

\begin{figure}
  \centering \includegraphics[width=90mm]{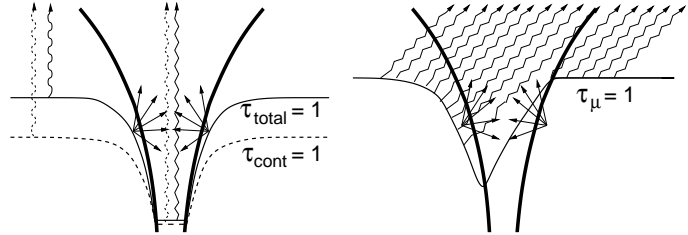}
  \caption[]{\label{rutten-fig:BPcartoon} %
    Hole-in-surface fluxtube brightening.  {\em Left\/}: radial
    viewing.  The magnetic-pressure evacuation deepens the photon
    escape layer to well below the outside surface and causes
    additional deepening for neutral-atom lines through ionization,
    for molecular lines through dissociation, and for strong-line
    wings through reduced damping.  The escape layers have similar
    temperatures inside and outside, but the tube has a flatter
    temperature gradient due to hot-wall irradiation.  {\em Right\/}:
    slanted near-limb viewing.  The tube evacuation causes facular
    sampling of the hot granule behind the tube, appearing as a bright
    stalk.  From \citet{Rutten1999d}, 
    after \citet{1976SoPh...50..269S}. 
  }
\end{figure}

\begin{figure}  
  \centering    
  \includegraphics[width=1.0\textwidth]{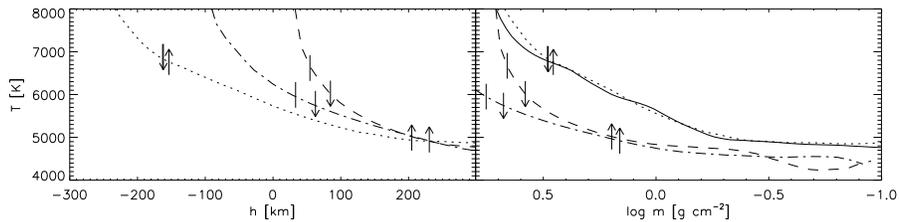}
  \caption[]{\label{rutten-fig:mni-models} Temperature stratifications
    in a fluxtube-like magnetic concentration (dotted), a granule
    (dashed) and an intergranular lane (dot-dashed) in a 3D MHD
    simulation.  {\em Left\/}: against geometrical height.  {\em
      Right\/}: against column mass.  {\em Markers\/}: $\tau \is 1$
    locations for \MnI\ 5394.7\,\AA\ (up arrows), \FeI\ 5395.2\,\AA\
    (down arrows), continuum in between (arrowless).  The solid curve
    is the PLA model of
    \citet{1992A&A...262L..29S} 
    derived from multi-line spectropolarimetry without angular
    resolution.  The fabulous agreement between this empirical best-fit
    model and the ab-initio simulation mutually vindicates these very
    different approaches.  Both spectral lines nearly vanish in the
    magnetic concentration from ionization.  The \MnI\ line is formed
    at much larger height in the granule and lane than the \FeI\ line
    which suffers weakening from convective Dopplershifts.  From
    \citet{Vitas+others2009}. 
  }\end{figure}

Also this old cartoon has been vindicated by recent simulations, for
continuum faculae by
\citet{2004ApJ...607L..59K}, 
for G-band faculae by
\citet{2004ApJ...610L.137C}, 
for bright points in the wings of strong lines by
\citet{2006A&A...452L..15L}, 
and for bright points in the unusually wide lines of \MnI\ by
\citet{Vitas+others2009}. 
The latter analysis, summarized in Fig.~\ref{rutten-fig:mni-models},
explains why the brightening seems less in all other, less wide,
photospheric lines: it is not, but they show the outside granulation
brighter from convective Dopplershift wash-out and so loose fluxtube
contrast.

These fluxtubes have near-radiative-equilibrium stratifications across
the upper photosphere, at least up to $h\approx400$~km as sampled by
the wings of \CaII\ \HK\
(\cite{2005A&A...437.1069S}), 
just as the mean outside photosphere above the granulation.  No
significant mechanical heating therefore: photospheric fluxtubes
brighten through deeper viewing.  All solar irradiance modeling
efforts that regard faculae as hot stick-up stalks and mimic such with
raised-temperature upper photospheres are inherently wrong (\eg\
\cite{1999A&A...345..635U}; 
\cite{2006ApJ...639..441F}; 
\cite{2007msfa.conf..189D}). 

The \NaDone\ image in Fig.~\ref{rutten-fig:na-ca-ha} suggests that
also the network brightening in this line is mostly hole-in-the-surface
viewing.  The outside $\tau\is1$ surface is formed in the upper
photosphere and very dark from scattering
(\cite{1992A&A...265..268U}), 
the tube inside is likely to be as bright as the continuum from
ionization.
This is now verifiable with Oslo STAGGER simulations.

The 5th panel of Fig.~\ref{rutten-fig:na-ca-ha} suggests that also
\CaII\ 8542\,\AA\ gains network brightening from surface-hole viewing
when not obscured by fibrils.  The STAGGER simulation of
\citet{2009ApJ...694L.128L} 
supports this notion.  The fluxtube at left in their Fig.~4 obtains
large brightness from deep line-center formation and Doppler
brightening in a post-shock downdraft, a likely occurrence in
deep-shocking fluxtubes
(\cite{1998ApJ...495..468S}). 

The \CaII\ \HK\ lines have larger opacity than \CaII\ 8542\,\AA\ in
any imaginable structure and reach larger opacity than \Halpha\ in
classical LTE
modeling\footnote{\url{http://www.astro.uu.nl/~rutten/rrweb/rjr-material/bachelor/ssa.}}
and in one-dimensional NLTE modeling
(\cite{1981ApJS...45..635V}). 
One would therefore expect to observe a denser fibril forest at right
in Fig.~\ref{rutten-fig:mosaics} than at left.  However, the DOT \CaIIH\ filter and all
other \HK\ filtergraphs have far too wide passbands to isolate
the line core, even the narrowest ones at 0.3\,\AA\
(\cite{1992ASPC...26..161B}; 
\cite{1998A&A...329..276H}; 
\cite{2006A&A...459L...9W}). 
In addition, fibrils are very dark in \HK\
(\cite{2008SoPh..251..533R}), 
so that the addition of bright structures in the inner wings dominates
the scene.  Magnetic concentrations are likely to again function as
hole-in-the-surface viewers, with the outside surface
clapotispherically cool and dark except in shocks, and the tube inside
and close surroundings bright by scattering photons from below (\cf\ Fig.~8
of \cite{Rutten1999d}, 
another old cartoon).

Thus, I contend that the apparent network brightness in \NaDone\ and
\CaII\ 8542\,\AA\ imaging spectroscopy and much of the apparent network
brightness in \CaII\ \HK\ filtergrams has less to do with
chromospheric heating than with hole-in-the-surface viewing.
Corollaries are that the Mount-Wilson \CaII\ \HK\ stellar activity
monitoring (\eg\ \cite{1986A&A...159..291R}; 
\cite{1991ApJS...76..383D}) 
delivers counts of such holes on the stellar surface, not the amount
of chromospheric heating, and that the good correspondence between the
various diagnostics of
\citet{2009A&A...497..273L} 
follows naturally because they are all hole counters.  And I repeat
that the basal flux (review:
\cite{1995A&ARv...6..181S}) 
is most likely clapotispheric (\cite{Rutten1999d}). 

\section{Heating in the chromosphere}  
So how to diagnose chromospheric heating, or rather, the onset of
coronal heating taking place in the chromosphere?  My contention that
most network emissivity in \NaDone, \CaII\ 8542\,\AA\ and \CaIIH\ in
Figs.~\ref{rutten-fig:mosaics} and \ref{rutten-fig:na-ca-ha} stems
from photospheric or clapotispheric view-pipe photon escape leads to
the suggestion that not much chromospheric heating is seen in these
images.  The \Halpha\ fibrils do display chromospheric fine-structure
morphology, but whether these appear dark or bright is a matter of
fortuitous combination of the four cloud parameters $S_\nu$,
$\tau_\nu$, $\lambda-\lambda_0$, $\Delta \lambda_{\rm D}$.  Only
$\Delta \lambda_{\rm D}$ senses the temperature directly because the
small atomic mass of hydrogen favors thermal over nonthermal
broadening.  The final panel of Fig.~\ref{rutten-fig:na-ca-ha}
therefore displays a measure of chromospheric temperature.  It
evidences heating in and near network
(\cite{Cauzzi++2009}). 

It seems likely that this network heating occurs through open-field
chromospheric structures, \ie\ more vertical than \Halpha\ fibrils and
connecting less closely than across network cells.  They are barely
visible in \Halpha\ but were discovered as fast-waving ``straws'' in
DOT \CaIIH\ near-limb movies, appearing bright against the very dark
internetwork clapotisphere
(\cite{2006ASPC..354..276R}), 
and subsequently as fast-waving ``spicules-II'' in Hinode \CaIIH\
off-limb movies by \citet{2007PASJ...59S.655D}. 
\citet{2007Sci...318.1574D} 
identified Alfv\'en waves as their modulation agent and estimated that
these accelerate the solar wind.  Being very slender and very
fast-moving, straws/spicules-II are very hard to observe on-disk
(\cite{2008ApJ...679L.167L}), 
but ongoing work (\cite{2008ESPM...12.2.15D}) 
suggests that they occur ubiquitously above unipolar network and plage
and may supply heating through component reconnection, reminiscent of
the separatrix shear proposed by Van Ballegooijen \etal\ (1998, 1999)
\nocite{1998ApJ...509..435V}
\nocite{1999ASPC..183...30V} 
and marked as cross-dashes in my old cartoon in
Fig.~\ref{rutten-fig:potsdam-f2}.  I suggest that unresolved straw
heating causes the network patches of large \Halpha\ core width in the
last panel of Fig.~\ref{rutten-fig:na-ca-ha}, some of the \CaII\
8542\,\AA\ near-network fibril brightness in the second panel of
Fig.~\ref{rutten-fig:na-ca-ha},
and the diffuse bright aureoles seen in DOT \CaIIH\ movies around
network that seem to resolve into straws in the very sharpest
one\footnote{
  \url{http://dot.astro.uu.nl/albums/movies/2006-04-24-NW-ca.avi} or
  \url{.mov}}.

Another speculation is that straw dynamics also contributes to the
solar-wind FIP fractionation.

\section{Discussion}  
There are striking and disconcerting similarities between the
visibility of the corona in the form of EUV-line loops and of the
chromosphere in the form of \Halpha\ fibrils.  Both types of structure
seem to map the large-scale magnetic field topology, although there is
no direct proof of direct correspondence.  In both cases it is a
misconception to regard them as magnetic fluxtubes embedded in
field-free plasma (or vacuum).  Loops and fibrils chart bundles of
field that for some reason contain more gas containing atoms and ions
in the right state for emissivity and/or extinction in the pertinent
lines than the surrounding, otherwise very similar field bundles.  The
actual field topology is probably more uniform than the fine-scaled
emissivity/extinction structuring suggests
(\cite{2006ASPC..354..259J}). 
The mass loading and dynamics of these narrow bundles are likely
structured on the scales of intergranular fluxtubes by thermodynamical
processes affecting these magnetic footpoints, but the actual connection
topology may be as complicated as in Fig.~18 of
\citet{2003ApJ...590..502D}. 

A disconcerting similarity of coronal loops and chromospheric fibrils
is that neither structure has yet been produced naturally in MHD
simulations.  Acoustic oscillations have been shown to drive and
probably mass-load short dynamic fibrils jutting out from network and
plage (\cite{2006ApJ...647L..73H}; 
\cite{2007ApJ...655..624D}), 
but the mass loading of the long network-spanning fibrils and yet
longer coronal loops remains unexplained.  The overpopulation arches
in Fig.~\ref{rutten-fig:hion-f1} are promisingly similar to actual
\Halpha\ fibrils, but they contain insufficient \Halpha\ opacity,
notwithstanding their enormous $n\is2$ overpopulation, to produce the
$\tau \!\approx\! 1$ thickness in \Halpha\ that is characteristic of
fibrils.  This simulation was 2D only, but also the newer 3D Oslo
STAGGER simulations have not produced fibrils or loops so far.
Perhaps the simulation volumes are still too small, or perhaps they
are too unipolar, or perhaps longer history is needed, perhaps to let
trickling siphon flows build up appreciable mass loads.

Another disconcerting similarity is that both coronal loops and
chromospheric fibrils delineate closed fields whereas the action lies
in the open-field components, respectively driving the fast wind and
heating the corona.  Their visibility is meager: coronal holes are
indeed holes in emissivity and chromospheric straws/spicules-II were
discovered only recently.  The natural emphasis on higher-visibility
loops and fibrils seems to target red herrings qua importance and
role.

Finally, it seems likely that the so-called transition region exists
mostly as hot evaporation sheaths around these ephemeral structures
(\cite{2008ApJ...673L.219M}; 
\cite{2008ApJ...687.1388J}; 
\cite{2009A&A...499..917K}), 
a far cry from the stable spherical shell invoked in one-dimensional
modeling or the radial stratification wishfully assumed in many 
transition-region oscillation propagation studies.

\begin{small}
\bibliographystyle{rr-assp}       

\end{small}

}
\end{document}